\documentclass{JHEP}
\usepackage{amsfonts}

\newcommand{\R}{\mathbb{R}}
\newcommand{\T}{\mathbb{T}}

\title{Comment on ``Casimir force in compact non-commutative
extra dimensions and radius stabilization''}
\author{Luiz C.\ de Albuquerque \\
Faculdade de Tecnologia de S\~ao Paulo - CEETEPS - UNESP \\
Pra\c{c}a Fernando Prestes, 30, 01124-060 S\~ao Paulo, 
SP, Brazil \\
E-mail: \email{lclaudio@fatecsp.br}}
\author{R.\ M.\ Cavalcanti \\
Instituto de F\'{\i}sica, Universidade de S\~ao Paulo \\
Caixa Postal 66318, 05315-970 S\~ao Paulo, SP, Brazil, and \\
Instituto de F\'{\i}sica, 
Universidade Federal do Rio de Janeiro \\
Caixa Postal 68528, 21945-970 Rio de Janeiro, RJ, 
Brazil\footnote{Present address} \\
E-mail: \email{rmoritz@if.ufrj.br}}

\abstract{We call attention to a series of mistakes
in a paper by S.\ Nam recently published in
this journal [\jhep{10}{2000}{044}].}


\preprint{}

\begin{document}

\setcounter{footnote}{0}

In a recent paper \cite{Nam}, S.\ Nam discussed
a possible mechanism for the stabilization of the
compact dimensions in a Kaluza-Klein scenario.
He considered the hypothesis of the geometry in the
compact dimensions being non-commutative, i.e., the
coordinates in these dimensions obeying the commutation
relations $\left[x^{\mu},x^{\nu}\right]=i\,\theta^{\mu\nu}$,
where $\theta^{\mu\nu}$ is an antisymmetric matrix.
This gives rise to the uncertainty relations
$\Delta x^{\mu}\,\Delta x^{\nu}\ge\frac{1}{2}\,|\theta^{\mu\nu}|$,
implying the existence of a minimum area in the $(\mu,\nu)$-plane,
which would prevent the collapse of the
compact dimensions.

In order to test this idea, Nam computed the one-loop
Casimir energy for the massless $\phi^3$ theory and the
$U(1)$ gauge theory defined on $\R^{1,d}\times \T_{\theta}^2$,
where $\R^{1,d}$ is an ordinary $(1+d)$-dimensional
Minkowski space-time and $\T_{\theta}^2$ is a non-commutative
two-torus whose coordinates satisfy $[y^1,y^2]=i\,\theta$
and $0\le y^1,y^2\le 2\pi R$. He arrived at the
conclusion that such a mechanism is not effective
in the scalar case, but it does work in the vector
case for $d>5$. Unfortunately he made a series of
mistakes which render his conclusions untenable.
The purpose of this Comment is to discuss those
mistakes.


\paragraph{First mistake.} In Ref.\ \cite{GMW}
Gomis {\it et al}.\ computed the one-loop correction to the
Kaluza-Klein spectrum for the $\phi^3$ theory and
the $U(1)$ gauge theory defined on $\R^{1,3}\times \T_{\theta}^2$.
Nam tacitly assumed that this correction is independent of the
dimension $d$ --- a wrong assumption, as Huang 
has shown \cite{Huang}.
As a consequence, Nam's results are not valid for $d\ne 3$.


\paragraph{Second mistake.} Even for $d=3$, Nam's calculation
of the Casimir energy density is incorrect. To begin with,
the sum in Eq.\ (13) of his paper is incomplete; 
it should run over 
all pairs of integers $(n_1,n_2)$ in the range
$-\infty<n_1,n_2<\infty$ (with the exception of the pair $n_1=n_2=0$
\cite{GMW}), instead of $n_1,n_2\ge 1$.
Therefore, the function $v_2$ in Eqs.\ (17)--(18) of Ref.\ \cite{Nam}
should be replaced by
\begin{equation}
\label{v2}
\tilde{v}_2(z)\equiv\sum_{m,n=-\infty}^{\infty}
\!\!\!\!^{'}\left(m^2+n^2\right)^{-z},
\end{equation}
the prime meaning that the term with $m=n=0$ is omitted.

The sum in Eq.\ (\ref{v2}) is convergent for ${\rm Re}\,z>1$,
so that Eq.\ (17) of \cite{Nam} is well defined for
$d<-3$. Before we can set $d=3$, we have to analytically continue
$\tilde{v}_2(z)$ to ${\rm Re}\,z<1$. To our purposes
it is enough to use the reflection formula \cite{Zucker}
\begin{equation}
\pi^{-z}\,\Gamma(z)\,\tilde{v}_2(z)=\pi^{z-1}\,\Gamma(1-z)\,
\tilde{v}_2(1-z),
\end{equation}
which allows us to rewrite Eq.\ (17) of \cite{Nam} as
[$\alpha\equiv\lambda^2/(4\pi)^3\theta^2$]
\begin{eqnarray}
\label{u'}
u(R)&=&-\frac{\mu^{3-d}}{2^{d+2}\,\pi^{(d+1)/2}}
\left[\pi^{-d-2}\,\Gamma\left(\frac{d+3}{2}\right)
\tilde{v}_2\left(\frac{d+3}{2}\right)R^{-d-1}\right.
\nonumber \\
& &\left.+\alpha\,\frac{2\pi^{-d+2}}{1-d}\,
\Gamma\left(\frac{d-1}{2}\right)
\tilde{v}_2\left(\frac{d-1}{2}\right)R^{-d+3}+\ldots\right],
\end{eqnarray}
which is now well defined for $d>3$.
(We have introduced an arbitrary mass scale $\mu$
to keep $u$ a three-dimensional energy density.)
Expanding $u$ in powers of $\epsilon=d-3$, we obtain
\begin{equation}
u(R)=-\frac{\tilde{v}_2(3)}{16\pi^7R^4}+\frac{\alpha}{16\pi^2}
\left(\frac{1}{\epsilon}-\ln(\mu R)+{\rm const}\right)+O(\epsilon).
\end{equation}
(To arrive at this expression, we have used the identity
$\tilde{v}_2(z)=4\,\zeta(z)\,\beta(z)$ \cite{Zucker},
where $\zeta(z)$ is the Riemann zeta function and
$\beta(z)\equiv\sum_{n=0}^{\infty}(-1)^n\,(2n+1)^{-z}$.)

The radiative correction to $u$ diverges when $d\to 3$.
However, the divergent contribution does not depend
on $R$, and so it can be subtracted from $u$, as it has no effect
on the Casimir force. In order to fix the finite part of
the correction, we impose that the renormalized Casimir
energy density vanishes at some arbitrary radius $R=R_*$; then
\begin{equation}
u(R)=-\frac{\tilde{v}_2(3)}{16\pi^7}\left(\frac{1}{R^4}
-\frac{1}{R_*^4}\right)-\frac{\alpha}{16\pi^2}\,
\ln\left(\frac{R}{R_*}\right).
\end{equation}
In contrast with the result found by Nam in the case $d=3$,
the above function has an extremum, located at
\begin{equation}
R=\left(\frac{4\,\tilde{v}_2(3)}{\pi^5\,\alpha}
\right)^{1/4}.
\end{equation}
This, however, does not imply the stabilization of the
compact dimensions, since this extremum is a
maximum of $u$. This leads us to Nam's


\paragraph{Third mistake.} Nam obtained for the $U(1)$
gauge theory on $\R^{1,d}\times\T_{\theta}^2$
a Casimir energy density of the form
\begin{equation}
u(R)=-\frac{a}{R^{d+1}}+\frac{b}{R^{d-5}},
\end{equation}
with $a$ and $b$ functions of $d$.
He then argued that
radius stabilization occurs for $d>5$, for in this
case the ratio $b/a$ is positive and there is a radius
$R_0$ such that $u'(R_0)=0$. However, he overlooked the
fact that $u''(R_0)<0$, so that the equilibrium is
{\em unstable}.


\paragraph{Fourth mistake.} Nam did not take into account all 
$O(\lambda^2)$ corrections to $u$.
Indeed, the Kaluza-Klein spectrum he used as the starting
point of his calculation takes into account only the
non-planar contribution to the one-loop self-energy \cite{GMW}.
However, the inclusion of the planar contribution is not
enough to correct the result. To show why, we start 
by noting that the
Casimir energy density can also be expressed 
as\footnote{While $u$ is a three-dimensional energy 
density, i.e.,
$u=E/{\rm Vol}(\R^3)$, $\epsilon$ is a five-dimensional
density: $\epsilon=E/{\rm Vol}(\R^3\times\T^2_{\theta})
=u/(2\pi R)^2$.} \cite{Fulling}
\begin{equation}
\label{e}
\epsilon = \langle 0_R|\,T_{00}\,|0_R\rangle,
\end{equation}
where $T_{00}$ is the 00-component of the 
energy-momentum tensor,
\begin{equation}
\label{T}
T_{00}=\frac{1}{2}\,(\partial_0\phi)^2+\frac{1}{2}\,
(\nabla\phi)^2+\frac{\lambda}{3!}\,\phi\star\phi\star\phi,
\end{equation}
and $|0_R\rangle$ is the vacuum state corresponding
to a compactification radius $R$. (Strictly speaking,
such a state does not exist in the present case,
as the Hamiltonian $H$ is unbounded from below.
In spite of this, one can still give a well defined 
meaning to Eq.\ (\ref{e})
by interpreting $|0_R\rangle$ as the state
which minimizes $\langle H\rangle$ under the
constraint $\langle\phi(x)\rangle =0$.)

Moving the derivatives outside the brackets, one can
rewrite Eq.\ (\ref{e}) as
\begin{equation}
\label{e2}
\epsilon(x)=\lim_{x'\to x}\,\frac{1}{2}\left(
\partial_0^{}\partial_0'+\partial_i^{}\partial_i'\right)
G^{(2)}(x,x')+\frac{\lambda}{3!}\,G^{(3)}_{\star}(x,x,x),
\end{equation}
where $G^{(2)}(x,x')\equiv\langle\phi(x)\phi(x')\rangle$ is the
(renormalized) connected two-point Green's function, and
$G^{(3)}_{\star}(x_1,x_2,x_3)\equiv\langle\phi(x_1)\star
\phi(x_2)\star\phi(x_3)\rangle$. The former satisfies the
Schwinger-Dyson equation,
\begin{equation}
\label{S-D}
-\partial^2G^{(2)}(x,x')
-\int dy\,\Sigma(x,y)\,G^{(2)}(y,x')=i\,\delta(x-x'),
\end{equation}
where $\Sigma$, the self-energy, is ($i$ times) the sum of all
1PI diagrams with two (amputated) external lines.
Using Eq.\ (\ref{S-D}) to eliminate the spatial derivatives
in Eq.\ (\ref{e2}), and noting that: (i) due to translation
invariance, $\partial_{\mu}'G^{(2)}(x,x')
=-\partial_{\mu}^{}G^{(2)}(x,x')$, and (ii)
$\lim_{x'\to x}\delta(x-x')=0$, we arrive at
\begin{equation}
\label{e3}
\epsilon(x)=\lim_{x'\to x}\,\partial_0^{}\partial_0'\,
G^{(2)}(x,x')-\frac{1}{2}\int dy\,\Sigma(x,y)\,G^{(2)}(y,x)
+\frac{\lambda}{3!}\,G^{(3)}_{\star}(x,x,x).
\end{equation}
Using the spectral representation of $G^{(2)}$, one can
recast the first term on the r.h.s.\ of Eq.\ (\ref{e3}) as
a sum of zero-point energies. Nam's calculation is 
equivalent to the computation of that term with 
$G^{(2)}$ approximated by
\begin{equation}
G^{(2)}(x,x')\approx i\,\langle x|\,\frac{1}
{-\partial^2-\Sigma^1_{\rm NP}}\,|x'\rangle ,
\end{equation}
where $\Sigma^1_{\rm NP}$ is the non-planar piece of the
one-loop self-energy. Hence, Nam's calculation includes
radiative corrections at  $O(\lambda^2)$. However,
besides neglecting the planar piece of
$\Sigma$, Nam overlooked the last two terms on the r.h.s.\
of Eq.\ (\ref{e3}), which also contain an $O(\lambda^2)$
contribution to the Casimir energy. 

In order to estimate the effect of those terms, let us consider,
for the sake of simplicity, the
commutative $\frac{\lambda}{3!}\,\phi^3$ theory.
Then, in the lowest non-trivial order in $\lambda$ the
self energy is given by
\begin{equation}
\Sigma(x,y)= -\frac{i\lambda^2}{2}
\left[G_0(x,y)\right]^2 +O(\lambda^4),
\end{equation}
where $G_0$ is the free two-point Green's function,
and the three-point Green's function reads
\begin{equation}
G^{(3)}(x,x,x)= -i\lambda\int dy
\left[G_0(x,y)\right]^3 +O(\lambda^3).
\end{equation}
The lowest order radiative correction to the Casimir 
energy density is thus given by
\begin{equation}
\label{e4}
\epsilon^{(2)}=\epsilon^{(2)}_{\rm Nam}+
\frac{i\lambda^2}{12}\int dy\left[G_0(x,y)\right]^3,
\end{equation}
where $\epsilon^{(2)}_{\rm Nam}$ is the result one obtains 
(in the commutative theory) by
following Nam's approach, i.e., by summing the one-loop
zero-point energies.

The integral in Eq.\ (\ref{e4}) can be evaluated
indirectly by noting that the Casimir energy density
[or, more precisely, $V_{\rm eff}(\langle\phi\rangle=0)$]
is also given by
\begin{equation}
\label{e5}
\epsilon=\lim_{V,T\to\infty}\frac{i}{VT}\,\ln Z,
\end{equation}
where
\begin{equation}
\label{Z}
Z=\int[{\cal D}\phi]_{\langle\phi\rangle=0}\,
\exp\left\{i\int dx\left[\frac{1}{2}\,(\partial\phi)^2
-\frac{\lambda}{3!}\,\phi^3\right]\right\}.
\end{equation}
(To satisfy the constraint $\langle\phi\rangle=0$,
one simply ignores any 1PI one-point diagrams in the
perturbative expansion of $Z$ \cite{Peskin}.)
In this approach, the second order term in the perturbative expansion
of $\epsilon$ is given by
\begin{equation}
\epsilon^{(2)}=-\frac{i\lambda^2}{12}
\int dy\left[G_0(x,y)\right]^3.
\end{equation}
Comparing with Eq.\ (\ref{e4}), we obtain
\begin{equation}
\epsilon^{(2)}_{\rm Nam}=-\frac{i\lambda^2}{6}
\int dy\left[G_0(x,y)\right]^3=2\,\epsilon^{(2)},
\end{equation}
i.e., the lowest order radiative correction to $\epsilon$
computed \`a la Nam is twice the correct value.
We expect a similar correction in the non-commutative theory. 
However, this has to be checked by an explicit computation.


\acknowledgments

This work was partially supported by FAPESP under grants
00/03277-3 (LCA) and 98/11646-7 (RMC), and by FAPERJ (RMC).
The authors acknowledge the kind hospitality of the Departamento de
F\'\i sica Matem\'atica, Universidade de S\~ao Paulo.


\end{document}